# The Mathematical Relationship between Zipf's Law and the Hierarchical Scaling Law

Yanguang Chen

(Department of Geography, College of Urban and Environmental Sciences, Peking University, 100871, Beijing, China. Email: chenyg@pku.edu.cn )

**Abstract:** The empirical studies of city-size distribution show that Zipf's law and the hierarchical scaling law are linked in many ways. The rank-size scaling and hierarchical scaling seem to be two different sides of the same coin, but their relationship has never been revealed by strict mathematical proof. In this paper, the Zipf's distribution of cities is abstracted as a $q$-sequence. Based on this sequence, a self-similar hierarchy consisting of many levels is defined and the numbers of cities in different levels form a geometric sequence. An exponential distribution of the average size of cities is derived from the hierarchy. Thus we have two exponential functions, from which follows a hierarchical scaling equation. The results can be statistically verified by simple mathematical experiments and observational data of cities. A theoretical foundation is then laid for the conversion from Zipf's law to the hierarchical scaling law, and the latter can show more information about city development than the former. Moreover, the self-similar hierarchy provides a new perspective for studying networks of cities as complex systems. A series of mathematical rules applied to cities such as the allometric growth law, the $2^n$ principle and Pareto's law can be associated with one another by the hierarchical organization.

# 1 Introduction

A network of cities is a complex system and can be treated as a hierarchy with cascaded structure. Hierarchy is frequently observed within the natural world as well as in human society.

The hierarchical structure always follows scaling laws and can be described with power functions. Equivalently, a power function can often be replaced by two exponential functions. A fractal, for example, is a typical hierarchy (Batty and Longley, 1994; Chen, 2009; Frankhauser, 1998). The hierarchical structure can be characterized by one power law or two exponential laws (Appendix 1). This mathematical description can be generalized to random fractals or statistical fractals, which are similar to cities. If hierarchical systems of cities are fractal, they can be described with a power law, such as Zipf's law and Pareto's law, or two exponential laws, such as Davis's $2^n$ rule (Chen and Zhou, 3003; Davis, 1978). The improvement of the $2^n$ principle of cities yields two general exponential models. A hierarchical power function can be derived from the two exponential functions. This suggests that a fractal hierarchy of cities can be modeled by employing two scaling laws: the rank-size scaling law, i.e., Zipf's law, and the hierarchical scaling law. However, the mathematical relationship between the two scaling laws has not yet been made clear. Revealing the relationship will bring to light much useful information of city development and present a new way of looking at complex systems.

Empirically, if the size distribution of cities in a region follows Zipf's law, the cities can be organized into a self-similar hierarchy composed of $M$ classes (levels) in a top-down order ($M$ is a positive integer) (Chen, 2012). The hierarchy of cities can be described with the discrete forms of two exponential functions such as

$$f_m = f_1 r_f^{m-1}, \tag{1}$$

$$P_m = P_1 r_p^{1-m}, \tag{2}$$

where $m=1, 2, \ldots, M$ refers to the order of classes in the hierarchy, $f_m$ and $P_m$ denote the city number and average city size in the $m$th class, $f_1$ and $P_1$ are the city number and size of the top-level cities, $r_f = f_{m+1}/f_m$ and $r_p = P_m/P_{m+1}$ are the number ratio and size ratio of cities, respectively. Equations (1) and (2) were empirically supported by the observational data of cities (Chen and Zhou, 2003). If $r_f=2$ as given, then the value of $r_p$ can be calculated (Chen, 2012); if $r_p=2$ as given, then the value of $r_f$ can be derived (Davis, 1978). Equations (1) and (2) express the generalized $2^n$ rule of cities (Chen and Zhou, 2003). From the two equations follows the hierarchical scaling law of cities in the form

$$f_m = \mu P_m^{-D}, \tag{3}$$

in which $\mu=f_1P_1^D$ refers to constant proportionality, and $D=\ln r_f/\ln r_p$ to the fractal dimension of urban hierarchies and city-size distributions (Chen, 2012).

In theory, equation (3) can be derived from Zipf's law. The association of the hierarchical scaling law with Zipf's law has been supported by many empirical studies of cities (Basu and Bandyapadhyay, 2009; Chen, 2009; Chen, 2012; Chen and Zhou, 2003; Jiang and Yao, 2010); However, there has never been a strict mathematical demonstration for it. In the next section, I will give mathematical proof for an equivalent relationship between Zipf's law and the hierarchical scaling law. The approach is as follows. First, the Zipf distribution is abstracted as a $q$-sequence. Second, the $q$-sequence is placed into a hierarchy according to equation (1). Third, based on the self-similar hierarchy, equation (2) will be proved to be true. Finally, equation (3) is derived from equations (1) and (2). The mathematical proof is so simple and elegant that the relationship between Zipf's law and the hierarchical scaling law becomes very clear.

The significance of this study lies in the fact that we can understand the Zipf distribution through the self-similar hierarchy. Many types of physical and social phenomena satisfy the well-known rank-size distribution and thus follow Zipf's law. The empirical law is named after the linguist Zipf (1949), but the scaling regularity of size distributions actually originates from Auerbach's urban studies in 1913 (Carroll, 1982). Today, Zipf's law has been used to describe the discrete power law probability distributions in various natural and human systems (Altmann *et al*, 2009; Axtell, 2001; Blasius and Tönjes, 2009; Brakman *et al*, 1999; Córdoba, 2008; Flam, 1995; Furusawa and Kaneko, 2003; Gabaix, 2009; Gabaix and Ioannides, 2004; Newman, 2005; Petersen *et al*, 2010; Podobnik *et al*, 2010; Serrano *et al*, 2009; Shao *et al*, 2007; Shao *et al*, 2011; Stanley *et al*, 1995). However, more information cannot be obtained from a system by using Zipf's law alone. On the other hand, many types of data associated with Zipf's law in the physical and social sciences can be arranged in a natural order to form a hierarchy with a cascade structure. By means of simple theoretical derivation, mathematical experiments, and empirical analyses, this paper will show that the Zipf distribution is equivalent to the self-similar hierarchical structure. As a typical example of rank-size distributions, American and Chinese cities are utilized to establish the observational foundation for this study.

## 2 Models and results

### 2.1 The self-similar hierarchy based on Zipf's law

Zipf's law is an empirical formulation which is used to describe the quantitative relationship between rank and size of cities in urban studies (Batty, 2008; Carroll, 1982; Gabaix and Ioannides, 2004; Marsili and Zhang, 1996). The 'law' states that, if the population of a city is multiplied by its rank to the power of $q$, the product will equal the population of the highest ranked city. In other words, the size of a city ranked $k$ will be $1/k^q$th of the size of the largest city. The general form of Zipf's law is

$$P_k = P_1 k^{-q}, \qquad (4)$$

where $P_k$ refers to the size of the city ranked $k$, and $P_1$ to the size of the largest city. If the Zipf exponent $q=1$, then equation (4) will change into the pure form of Zipf's law which indicates the so-called *rank-size rule* (Knox and Marston, 2006). If $P_1=1$ unit, then the city size can be abstracted as a $q$-sequence $\{1/k^q\}$, where $k=1, 2, 3, …$, (a sequence of natural numbers) .

The $q$-sequence is the generalized result of the harmonic series. The sequence can be made into a self-similar hierarchy by rearranging it according to the geometric proportion rule. For simplicity, we can construct the hierarchy in terms of equation (1). Suppose that different classes are defined as different levels of a hierarchy. One fraction for the first class, two fractions for the second class, four fractions for the third class, and so on. Generally, we have $2^{m-1}$ fractions for the $m$th class ($m=1, 2, 3, …$). The first four classes can be expressed as follows: [$1/1^q$], [$1/2^q$, $1/3^q$], [$1/4^q$, $1/5^q$; $1/6^q$, $1/7^q$], [$1/8^q$, $1/9^q$; $1/10^q$, $1/11^q$; $1/12^q$, $1/13^q$; $1/14^q$, $1/15^q$]. In this way, the $q$-sequence $\{1/k^q\}$ is linked with the geometric sequence $\{2^i\}$, where $i=0, 1, 2, 3, …..$ The pattern of construction is illustrated in Figure 1, which shows the doubling effect of the self-organizing systems of cities.

The self-similar hierarchy displayed in Figure 1 possesses a stable structure. At the micro-level, as shown by the Batty's rank clocks, cities rise and fall in size at many times and on many scales (Batty, 2006). However, at the macro-level, the hierarchical structure changes little. This brings to mind the 'hidden order' pointed out by Holland (1995, page 1), who observed: "Like the standing wave in front of a rock in a fast-moving stream, a city is a pattern in time. No single constituent remains in place, but the city persists." In fact, the Zipf distribution is a typical pattern in time. No single city remains in the same rank, but the rank-size pattern persists. It is very easy to formulate this hierarchical structure. The number of cities in the $m$th class, $f_m$, can be defined as

$$f_m = 2^{m-1} = f_1 r_f^{m-1}, \tag{5}$$

where $f_1=1$ denotes the number of elements in the top-ranked class. This is just equation (1), which can be termed the *number law* of urban hierarchies. A common ratio of interclass, the above-mentioned number ratio, is as follows

$$r_f = \frac{f_{m+1}}{f_m} = 2. \tag{6}$$

Now, the key to proving the relationship between Zipf's law and the hierarchical scaling law is to derive equation (2) from the hierarchy. If it can be proved that the sum of each class, $S_m$, approaches a geometric sequence, then the mean of each class will also form a geometric sequence, and satisfies the exponential distribution. In this instance, the common ratio will approach without limit to $1/2^q$, and thus the size ratio of different classes of cities is infinitely close to $2^q$. As a consequence, the hierarchy of cities takes on the standard cascade structure, which can be described using a discrete power law or two exponential laws, such as equations (1), (2), and (3).

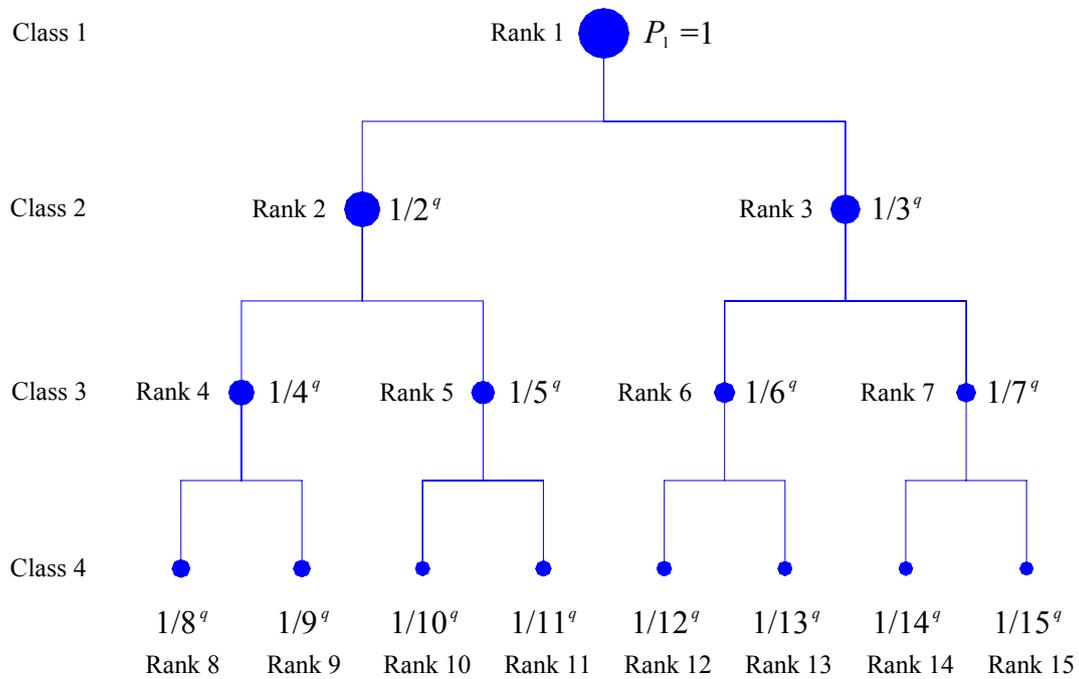

**Figure1** The schematic diagram of the self-similar hierarchy of cities based on Zipf's law and the generalized $2^n$ rule

## 2.2 Derivation of the generalized $2^n$ rule from Zipf's law

The equivalence relation between Zipf's and the hierarchical scaling law can be formulated as a simple problem of mathematical proof. According to the structure of the self-similar hierarchy of $q$-sequence (Figure 1), the summation of numerical values in each level, $S_m$, can be expressed as

$$S_m = \frac{1}{(2^{m-1})^q} + \frac{1}{(2^{m-1}+1)^q} + \cdots + \frac{1}{(2^m-1)^q} = \sum_{j=0}^{2^{m-1}-1} \frac{1}{(2^{m-1}+j)^q}, \quad (7)$$

in which $j=0, 1, 2, \cdots, 2^{m-1}-1$. The problem of Zipf's law being equivalent to the hierarchical scaling law will be resolved if we can prove the following relation

$$\lim_{m \to \infty} S_m \propto 2^{(1-q)(m-1)}. \quad (8)$$

In fact, equation (7) can be equivalently rewritten in the following form

$$S_m = \frac{1}{\left[2^{m-1}(1+\frac{0}{2^{m+1}})\right]^q} + \frac{1}{\left[2^{m-1}(1+\frac{1}{2^{m+1}})\right]^q} + \cdots + \frac{1}{\left[2^{m-1}(1+\frac{2^{m-1}-1}{2^{m-1}})\right]^q}$$

$$= \sum_{j=0}^{2^{m-1}-1} \left(\frac{\frac{1}{2^{m-1}}}{1+\frac{j}{2^{m-1}}}\right)^q = \left(\frac{1}{2^{m-1}}\right)^{q-1} \cdot \sum_{j=0}^{2^{m-1}-1} \frac{\frac{1}{2^{m-1}}}{(1+\frac{j}{2^{m-1}})^q}. \quad (9)$$

Suppose that the function $f(x)=1/(1+x)$ is defined within the closed interval $[0, 1]$. Dividing the interval into $2^{m-1}$ equal parts by using the fractions

$$0 = \frac{0}{2^{m-1}}, \frac{1}{2^{m-1}}, \frac{2}{2^{m-1}}, \cdots, \frac{2^{m-1}-1}{2^{m-1}}, \frac{2^{m-1}}{2^{m-1}} = 1.$$

The length of each part is just $1/2^{m-1}$. In terms of the attribute of $j$, let

$$x_j = \frac{j}{2^{m-1}}. \quad (10)$$

Thus we have

$$\Delta x_j = \frac{1}{2^{m-1}}. \quad (11)$$

Moreover, if $m$ is large enough, then

$$\max(x_j) = \lim_{m \to \infty} \frac{2^{m-1}-1}{2^{m-1}} = 1. \quad (12)$$

This suggests that $x_j \in [0,1]$, corresponding to the interval divided by the above fractions. If $q \neq 1$,

the summation of each class is as follows

$$S_m = (2^{m-1})^{1-q} \lim_{m\to\infty} \sum_{j=0}^{2^{m-1}-1} \frac{\frac{1}{2^{m-1}}}{(1+\frac{j}{2^{m-1}})^q} = 2^{(1-q)(m-1)} \lim_{m\to\infty} \sum_{j=0}^{2^{m-1}-1} \frac{\Delta x_j}{(1+x)^q}, \quad (13)$$

$$\to 2^{(1-q)(m-1)} \int_0^1 \frac{1}{(1+x)^q} dx = \frac{2^{1-q}-1}{1-q} 2^{(1-q)(m-1)} = S_1 2^{(1-q)(m-1)}$$

where $S_1=(2^{1-q}-1)/(1-q)$, and the arrow '→' means 'to be infinitely close to'. The proof of equation (8) is complete, and the conclusion can be reached that the sum of numerical values in each class does approach a geometric sequence.

Next, the average size of each class can be demonstrated to follow the exponential law. Dividing equation (13) by equation (5) yields equation (2) such as

$$P_m = \frac{S_m}{f_m} = \frac{2^{1-q}-1}{1-q} 2^{q(1-m)} = P_1(2^q)^{1-m} = P_1 r_p^{1-m}, \quad (14)$$

where $P_m$ is the average size of the $m$th class, the parameter

$$P_1 = \frac{S_1}{f_1} = \frac{2^{1-q}-1}{1-q} \le 1 \quad (15)$$

refers to a proportionality coefficient. If $f_1=1$, then theoretically $P_1=S_1$. Equation (14) denotes the *size law* of urban hierarchies. The common ratio of the sequence of average size is

$$r_p = \lim_{m\to\infty} \frac{P_m}{P_{m+1}} = 2^q. \quad (16)$$

The fractal dimension of the hierarchy of cities can be defined by

$$D = \frac{\ln r_f}{\ln r_p} \to \frac{\ln(2)}{\ln(2^q)} = \frac{1}{q}, \quad (17)$$

Thus, the fractal property of the Zipf distribution has been demonstrated (Mandelbrot, 1983; Frankhauser, 1990; Makse *et al*, 1995; Makse *et al*, 1998; Rozenfeld *et al*, 2011). Equations (1) and (2), or equations (5) and (14) illustrate two exponential laws known as the general $2^n$ principle of cities (Chen and Zhou, 2003). It is not difficult to generalize the rule to the well-known Horton-Strahler's laws (Horton, 1945; Strahler, 1952), Gutenberg-Richter's laws (Gutenberg and Richter, 1954), and other such laws.

The pure form of Zipf's law should be clarified, and it will be specifically discussed in a companion paper. According to l'Hospital's rule, if $q=1$, then

$$S_1 = \lim_{q \to 1} \frac{2^{1-q}-1}{1-q} = \frac{2^{1-q}\ln(2)\mathrm{d}(1-q)}{\mathrm{d}(1-q)} = \ln(2). \tag{18}$$

Thus

$$\lim_{m \to \infty} S_m = S_1 = \ln(2), \tag{19}$$

and

$$P_m = \ln(2) \cdot 2^{1-m}. \tag{20}$$

This indicates that, if $q=1$, then $P_1=\ln(2)$ and $r_p=2$. Therefore, in light of equation (5) and equation (20), the product of city number and average city size of each class is $f_m P_m=\ln(2)$. In this case, the fractal dimension of the urban hierarchy equals $D=1/q=1$, which implies the standard $2^n$ rule proposed by Davis (1978).

## 2.3 Scaling reconstruction of Zipf's law

So far, the main work of mathematical proof has been completed. An inverse power-law relationship between average city size and the number of cities at each level can be readily derived from equations (5) and (14), and the result is just equation (3). Thus the rank-size scaling law, equation (4), can be replaced by the size-number scaling law, equation (3). In practice, when $m$ is too small or the class is too high, the data points of city size v.s. number probably get departure to some extent from the scaling range on the log-log plots. Theoretically, if $m=1$ as given, then $\mu=f_1 P_1^D=1>S_1=(2^{1-q}-1)/(1-q)$. The $q$ value is between 0.5 and 2 (Chen, 2011). So the $S_1$ value varies from 0.5 to 0.828. However, the result of mathematical derivation is $\mu=(2^{1-q}-1)/(1-q)$ rather than 1. This suggests that the fractal structure of the hierarchy of cities based on the $q$-sequence comes into existence asymptotically when $m$ increases. This is consistent with the definition of fractal dimension, which is a parameter defined under limit condition (Mandelbrot, 1983). It can be inferred that the Zipf distribution is asymptotically equivalent to the self-similar hierarchical structure as $m$ approaches a large number. As for the empirical data, the last class is always beyond the scaling range because of slower than expected growth in small cities and towns. Therefore, the power laws of urban hierarchies are usually invalid at the extreme edges of the scales, i.e. the very large and small scales. The conclusion is consistent with that from the entropy-maximizing analysis (Chen, 2012).

# 3 Materials and methods

## 3.1 Mathematical experiments

The equivalence of Zipf's law to the hierarchical scaling law can be visually understood by simple mathematical experiments. A mathematical experiment can be defined as an approach to using computational methods based on certain models or postulates to verify a mathematical law or inference. According to Zipf's law, if $P_1=1$ as given, then the various city sizes will make a $q$-sequence, $[1, 1/2^q, 1/3^q, \ldots, 1/k^q, \ldots]$, which can be placed into a hierarchy with a cascade structure, as displayed in Figure 1. For example, there is one city, which is the first city by rank, in the first class; two cities, the second and third one by rank, in the second class; four cities, the fourth to the seventh one, in the third class, and so on. The sequence in the $m$th class is as follows: $[1/(2^{m-1})^q, 1/(2^{m-1}+1)^q,\ldots,1/(2^m-1)^q]$. Clearly, the numbers of cities in different levels form a geometric sequence, which can be perfectly fitted into equation (1), and the common ratio is $r_f=2$, where $m=1, 2, \ldots, M$.

Now, let's see the sum of numerical values in each class. In the first class, the sum is $S_1=1$; in the second class, the sum is $S_2=1/2^q+1/3^q$; in the third class, the sum is $S_3=1/4^q+1/5^q+1/6^q+1/7^q$, and so on. If $q=1$ as given, then the sum $S_m$ approaches $\ln(2)$ rapidly when $M$ increases. Thus the mean value of each class quickly approaches an exponential distribution such as $P_m=\ln(2)/(2^{m-1})$. As a result, the average sizes constitute a geometric sequence with a common ratio 1/2. This suggests $r_p=P_m/P_{m+1}\to f_{m+1}/f_m=2=r_f$. In this instance, the fractal dimension $D=\ln r_f/\ln r_p=1/q\to 1$. If $q<1$, the sum, $S_m$, is equivalent to a positive exponential function (geometric growth); while if $q>1$, $S_m$ can be fitted to a negative exponential function (geometric decay). Anyway, the average size, $P_m$, exhibits an exponential decay. The $q$ value proved to range from 0.5 to 2 (Chen, 2011). Partial results of typical mathematical experiments are listed in Table 1 by taking $q=0.8$, $q=1$, and $q=1.2$ ($M=10$). The regularity is evident: when $m$ goes up, the average size goes down, and the size ratio becomes closer and closer to the constant $r_p=2^q$.

Next, we can examine the fractal dimension value of urban hierarchies. Since cities satisfy the standard rank-size distribution, the fractal dimension $D=1/q$ is a known number. If Zipf's law is equivalent to the hierarchical scaling law, the estimated value of fractal dimension through the hierarchical scaling should also be $1/q$ or close to $1/q$. According to equation (17), if $m \to \infty$, then $D \to 1/q$. Based on these mathematical experiments, Table 1 shows that the equivalence relation between Zipf's law and the hierarchical scaling law is true in theory. In reality, the number of classes/levels of urban hierarchies is limited, but when $m > 1/\ln(r_p)$, where $1/\ln(r_p)$ is the characteristic length of the exponential distribution, we can obtain good approximate results that support the theoretical inference.

Table 1 The results of mathematical experiments by converting the $q$-sequences based on Zipf's law into geometric sequences based on the generalized $2^n$ rule ($r_f=2$)

| Parameter ($q$) | Class ($m$) | City number ($f_m$) | Total city number ($f_m P_m$) | Average size ($P_m$) | Size ratio ($r_p$) |
|---|---|---|---|---|---|
| $q=0.8$ $D=5/4$ | 1 | 1 | 1 | 1 | |
| | 2 | 2 | 0.9896 | 0.4948 | 2.0210 |
| | 3 | 4 | 1.0551 | 0.2638 | 1.8758 |
| | 4 | 8 | 1.1684 | 0.1460 | 1.8062 |
| | 5 | 16 | 1.3180 | 0.0824 | 1.7730 |
| | 6 | 32 | 1.5004 | 0.0469 | 1.7569 |
| | 7 | 64 | 1.7158 | 0.0268 | 1.7489 |
| | 8 | 128 | 1.9665 | 0.0154 | 1.7450 |
| | 9 | 256 | 2.2564 | 0.0088 | 1.7431 |
| | 10 | 512 | 2.5904 | 0.0051 | 1.7421 |
| | … | … | … | … | … |
| | $M$ | $2^{M-1}$ | $S_M$ | $P_1/[(2^{0.8})^{M-1}]$ | $2^{0.8}$ |
| $q=1$ $D=1$ | 1 | 1 | 1 | 1 | |
| | 2 | 2 | 0.8333 | 0.4167 | 2.4000 |
| | 3 | 4 | 0.7595 | 0.1899 | 2.1944 |
| | 4 | 8 | 0.7254 | 0.0907 | 2.0942 |
| | 5 | 16 | 0.7090 | 0.0443 | 2.0461 |
| | 6 | 32 | 0.7010 | 0.0219 | 2.0228 |
| | 7 | 64 | 0.6971 | 0.0109 | 2.0113 |
| | 8 | 128 | 0.6951 | 0.0054 | 2.0057 |
| | 9 | 256 | 0.6941 | 0.0027 | 2.0028 |
| | 10 | 512 | 0.6936 | 0.0014 | 2.0014 |
| | … | … | … | … | … |
| | $M$ | $2^{M-1}$ | $\ln(2)$ | $\ln(2)/(2^{M-1})$ | 2 |
| $q=1.2$ $D=5/6$ | 1 | 1 | 1 | 1 | |
| | 2 | 2 | 0.7029 | 0.3514 | 2.8455 |
| | 3 | 4 | 0.5477 | 0.1369 | 2.5666 |
| | 4 | 8 | 0.4511 | 0.0564 | 2.4282 |
| | 5 | 16 | 0.3821 | 0.0239 | 2.3615 |
| | 6 | 32 | 0.3281 | 0.0103 | 2.3291 |

| | | | | |
|---|---|---|---|---|
| 7 | 64 | 0.2837 | 0.0044 | 2.3132 |
| 8 | 128 | 0.2461 | 0.0019 | 2.3053 |
| 9 | 256 | 0.2139 | 0.0008 | 2.3013 |
| 10 | 512 | 0.1860 | 0.0004 | 2.2994 |
| … | … | … | … | … |
| M | $2^{M-1}$ | $S_M$ | $P_1/[(2^{1.2})^{M-1}]$ | $2^{1.2}$ |

However, for the empirical analysis of observational data, the computation is based on statistical averages. Therefore, a statistical approach to data processing is presented as follows. Step 1: compute the size ratio ($r_p$) with the formula $r_p=P_m/P_{m+1}$. If $M=10$ as given, then there will be nine numbers of $r_p$, which will approach $2^q$ gradually. Step 2: standardize the sequence of size ratios by means of the z-score formula. Step 3: determine the scaling range by plotting the data on a log-log graph. Theoretically, the size ratio will be constant. If one or two standardized data are greater than the double standard error 2, these data should be removed as outliers. Generally speaking, the outliers appear at the beginning or end of a sequence. The calculations show that only the first ratio is an exception. Therefore, the first and the second class can be regarded as the points beyond the scaling range. A linear regression of $\ln f_m$ on $\ln P_m$ can be implemented by using data from the third class to the tenth class. The scaling relations are displayed in Figure 2 for $q=0.8$, $q=1$, and $q=1.2$. For example, if $q=1$, the least squares computation involving the data of scaling range yields the following model:

$$\hat{f}_m = 0.757 P_m^{-0.984},$$

where the hat '^' implies that the result is of estimation. The goodness of fit is approximately $R^2=0.9999$, and the fractal dimension is estimated as $\hat{D} \approx 0.984$. If $M \gg 10$, then $\hat{D} \to 1$, and accordingly, $R^2 \to 1$, suggesting the perfect fit. Changing $q$ values yields different results for each mathematical experiment. Each $q$ value corresponds to a fractal dimension $D=1/q$. The fractal dimension estimated from urban hierarchical structure ($\hat{D}$) is very close to the prearranged value from the rank-size distribution ($D$), as shown in Table 2.

**Table 2** Theoretical expectation ($D$) and empirical results ($\hat{D}$) of the fractal dimension corresponding to different $q$ values of the Zipf distribution ($M=10$)

| q | 0.200 | 0.400 | 0.600 | 0.800 | 1.000 | 1.200 | 1.400 | 1.600 | 1.800 | 2.000 |
|---|---|---|---|---|---|---|---|---|---|---|
| D | 5.000 | 2.500 | 1.667 | 1.250 | 1.000 | 0.833 | 0.714 | 0.625 | 0.556 | 0.500 |

| $\hat{D}$ | 4.923 | 2.462 | 1.641 | 1.231 | 0.984 | 0.820 | 0.703 | 0.615 | 0.547 | 0.492 |

**Note:** For each $q$ value of standard $q$-sequence of cities, there is a theoretical expectation of fractal dimension $D=1/q$. The dimension value $D$ can be estimated by the formula $D=\ln r_f/\ln r_p$, and the result is notated as $\hat{D}$. If $M$ is large enough, the $D$ value will be very close to the $\hat{D}$ value.

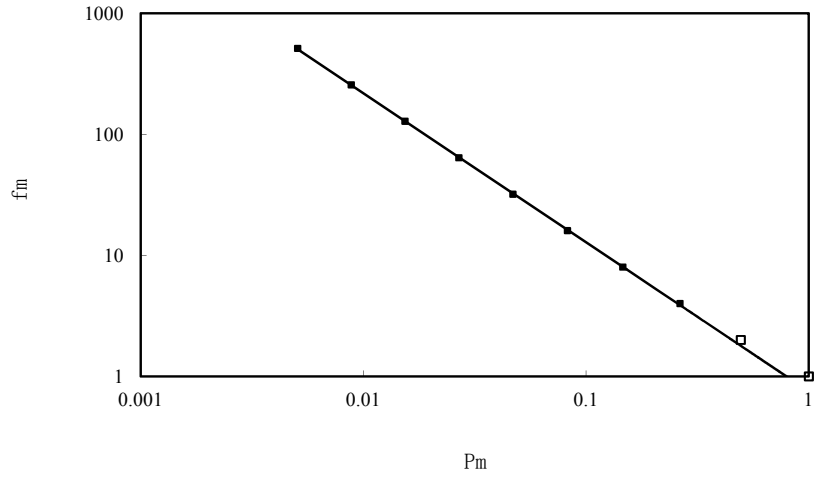

a. $q=0.8$, $D=1.25$

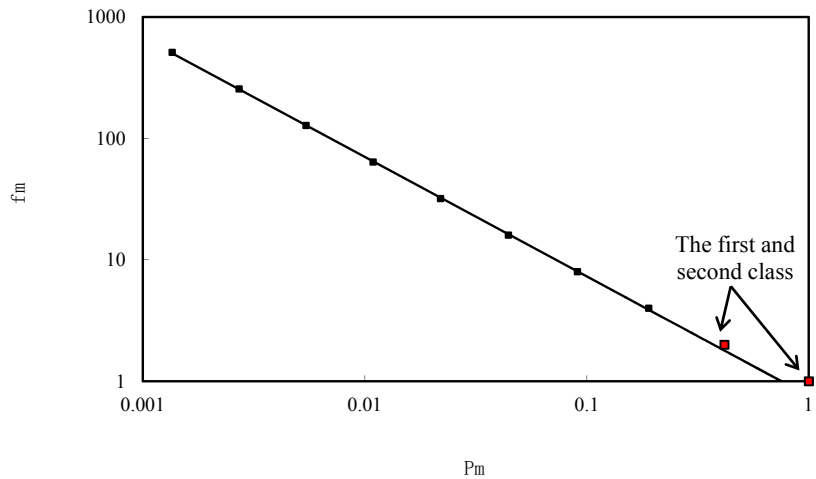

b. $q=1$, $D=1$

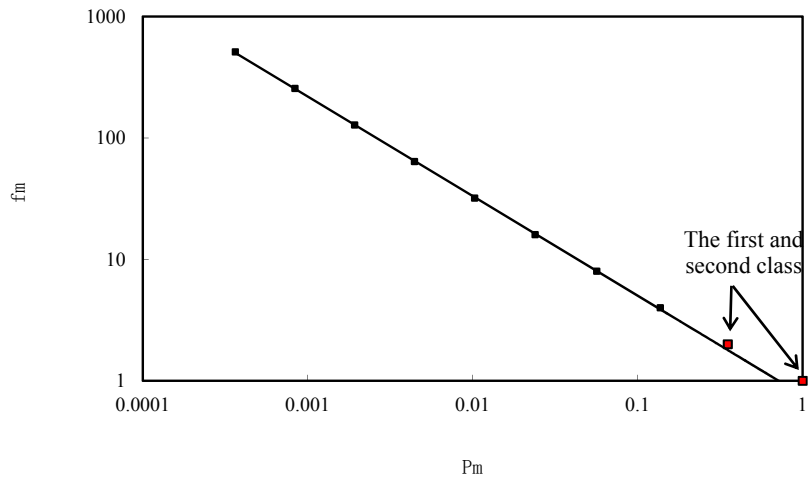

c. *q*=1.2, *D*=0.833

**Figure 2** The hierarchical scaling relation between city numbers and average sizes of different classes (**Note:** According to the fact that the standard error of $r_p$ is greater than 2, the first and second classes are treated as exceptional values. The hollow squares represent the special data points beyond the scaling range.)

For the sake of simplicity, only the first 10 classes are taken into account, so the results are approximate. The number of classes *M* can theoretically approach infinity and can be large enough in practice. The greater the value of *M*, the closer the result will be to the theoretical expectation. For cities in the real world, the scaling invariance of urban hierarchies is only seen in a certain range of scales. If the scale is too large (corresponding to a small *m* value), the size ratio may not be fixed; if the scale is too small (corresponding to a large *m* value), the city number will always be insufficient due to undergrowth of small cities and towns (Chen, 2012).

### 3.2 Empirical evidence

The mathematical relationship between Zipf's law and the hierarchical scaling law can be further verified with observational data. The rules followed by the cities' distribution are all based on the notion of statistical averages, and the regularity of urban development is often perceived only on a large scale (Manrubia S, Zanette, 1998; Zanette and Manrubia, 1997). Precisely because of this, the cities in the United States of America (USA) and those of the People's Republic of China (PRC) are taken as two examples for the case studies. Generally, urban area and population can be defined by four basic concepts: *city proper* (CP), *urbanized area* or *urban agglomeration* (UA), and *metropolitan area* (MA). Different definitions of cities affect the statistical distribution of urban activity (Gabaix, 1999; Rozenfeld *et al*, 2008), but for Zipf's law, definitions of city mainly influence the scaling exponent values rather than the power-law relationship (Chen, 2011; Chen, 2012). The datasets of 601 U.S. cities and 666 P.R.C. cities in 2000 are available from a website (see the note below Table 3). In this paper, the U.S. city size is measured by population in the city proper, and the Chinese city size is measured by population in urban agglomeration. Both the American cities and the Chinese cities satisfy the Zipf distribution (Figure 3). Next, it will be determined whether or not the two sets of cities follow the hierarchical scaling law. In practice, there are two ways to approach this type of empirical analyses. One is to compose the self-similar hierarchy in terms of equation (1), and then fit the average sizes of different classes into equation

(2) (Chen, 2012); the other is to compose the hierarchy in terms of equation (2), and then fit the city numbers of different classes into equation (1) (Chen and Zhou, 2003; Davis, 1978). For convenience, I will adopt the first approach in this case study.

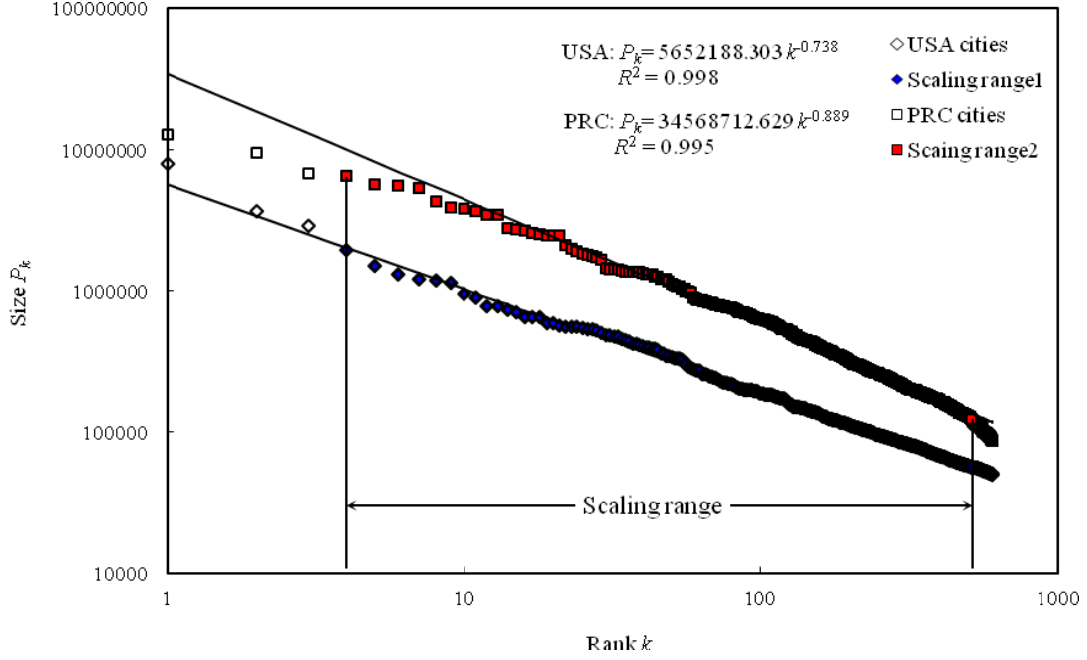

**Figure 3** The rank-size distributions of the USA cities and PRC cities in 2000
(**Note:** The scaling range is determined by the hierarchies of America's cities and China's cities. The first two classes including 3 cities and the last class including 90 cities and 155 cities respectively are treated as outliers.)

The first case involves the 601 U.S. municipalities with population size over 50,000 in 2000. Taking number ratio $r_f=2$, we can organize these cities in a hierarchical system. There is one city, New York, in the first class, two cities, Los Angeles and Chicago, in the second class, four cities, Houston, Philadelphia, Phoenix, and San Diego, in the third class, and eight in the fourth class, and so on. The rest may be addressed by analogy. That is, the city numbers in different levels can be formulated as

$$f_m = 2^{m-1} \approx 0.5 e^{-0.693m}.$$

If the average city size satisfies the exponential distribution, equations (1) and (2) will be supported and give equation (3). The city number, total urban population, and average population size in each level are listed in Table 3. Fitting the city size data to equation (2) yields

$$\hat{P}_m = 7568976.650 e^{-0.511m} = 4538873.590 \times 1.668^{1-m}.$$

The goodness of fit is about $R^2=0.994$, and the fractal dimension of the urban hierarchy is estimated as $D≈0.693/0.511≈1.356$ (See Appendix 2). Of course, the first data point is treated as an outlier because the extreme scale is always exceptional for the exponential distribution (Clark, 1951).

If we employ equation (3) directly to make a scaling analysis, the effect of fractal dimension estimation is better, but we must take into account the scaling range. The city number of the 10th class is expected to be $2^9=512$, but there is only data for 90 cities available. Therefore, the bottom level is in fact a 'lame duck class' according to Davis (1978). The size ratio changes by around 2 and fluctuates between 1.34 and 2.43. The double logarithmic plot of city number $f_m$ vs average city size $P_m$, which is displayed in Figure 4, shows that the first two classes and the last class can be treated as exceptional values when compared with the results of mathematical experiments shown in Figure 2. The rest, from the 3rd to the 9th class, falls into the scaling range of the hierarchical power law.

Let's compare the scaling exponent value from equation (4) with that from equation (3). The expected result is that the two values are very close to each other. A least squares computation involving the data points within the scaling range gives

$$\hat{f}_m = 1103924432.988 P_m^{-1.364}.$$

The correlation coefficient square is $R^2≈0.999$, and the fractal dimension of the hierarchy of America's cities is estimated as $D≈1.364$. Correspondingly, a least squares computation based on Zipf's law with the data of cities ranked 3 to 511 (see Figure 3), which is consistent with the scaling range of the urban hierarchy shown in Figure 4(a), yields a model such as

$$\hat{P}_k = 5652188.303 k^{-0.738}.$$

The correlation coefficient square is $R^2≈0.998$, and the fractal dimension of city-size distribution is $D=1/q≈1/0.738≈1.355$, close to the value 1.364 (the standard error $δ≈0.021$). The fractal dimension is approximately $D=4/3$, while the Zipf exponent is about $q=3/4$, which brings to mind the well-known 3/4 power law (West et al, 1999; West et al, 2002).

**Table 3** The results of empirical analyses by grouping America's cities and China's cities in 2000 into self-similar hierarchies according to the generalized $2^n$ rule ($r_f=2$)

| Class | USA cities | | | | PRC cities | | | |
| --- | --- | --- | --- | --- | --- | --- | --- | --- |
| (*m*) | $f_m$ | $S_m$ | $P_m$ | $r_p$ | $f_m$ | $S_m$ | $P_m$ | $r_p$ |
| 1 | 1 | 8008278 | 8008278 | | 1 | 12720701 | 12720701 | |
| 2 | 2 | 6590836 | 3295418 | *2.430* | 2 | 16365424 | 8182712 | *1.555* |
| 3 | 4 | 6015626 | 1503907 | 2.191 | 4 | 23115494 | 5778874 | 1.416 |
| 4 | 8 | 7185129 | 898141 | 1.674 | 8 | 28107178 | 3513397 | 1.645 |
| 5 | 16 | 9025334 | 564083 | 1.592 | 16 | 32977845 | 2061115 | 1.705 |
| 6 | 32 | 11587949 | 362123 | 1.558 | 32 | 37442973 | 1170093 | 1.761 |
| 7 | 64 | 12743202 | 199113 | 1.819 | 64 | 42724711 | 667574 | 1.753 |
| 8 | 128 | 15195535 | 118715 | 1.677 | 128 | 44348064 | 346469 | 1.927 |
| 9 | 256 | 18421117 | 71957 | 1.650 | 256 | 45323701 | 177046 | 1.957 |
| 10 | 90 | 4823966 | 53600 | *1.342* | 155 | 13355373 | 86164 | *2.055* |

**Note**: The original data from the 2000 census of the U.S. cities are available from: http://www.demographia.com, and the data of the P.R.C. cities from the 5th census of China are available from: http://pdfdown.edu.cnki.net.

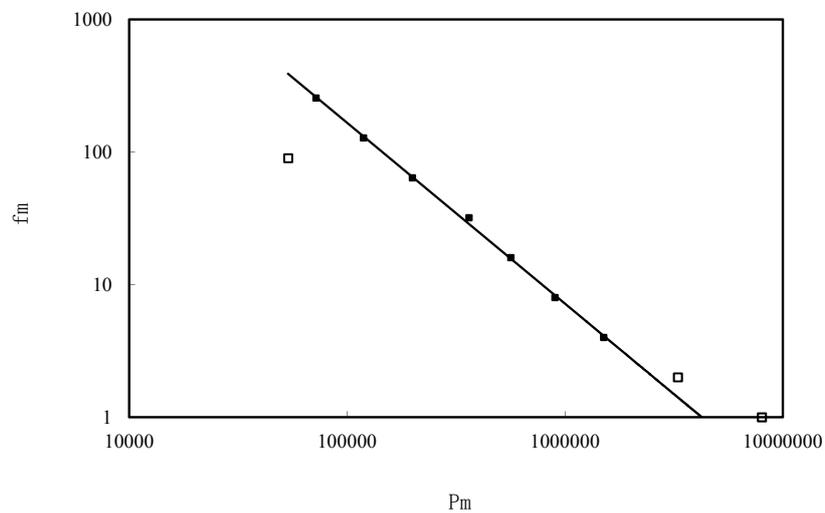

a. USA cities

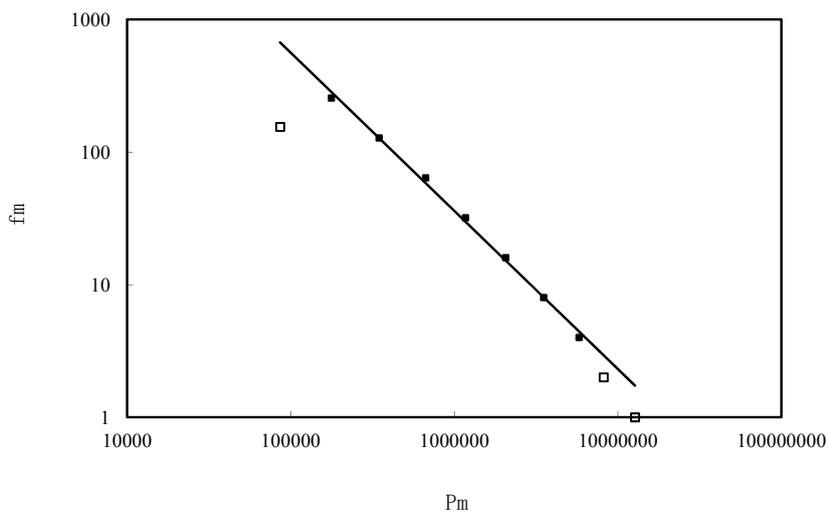

b. PRC cities

**Figure 4** The hierarchical scaling relation between city numbers and average sizes in the hierarchies of

USA and PRC cities in 2000

(**Note:** According to the results of mathematical experiments, the first and second classes are treated as exceptional values. Because of undergrowth of cities, the last classes are treated as outliers. The hollow squares represent the special data points beyond the scaling range.)

The same method of data processing and analysis is then applied to China's 666 cities in 2000. The result is similar to that of America. The size distribution of cities is rearranged by putting the largest city (Shanghai) in the first class, then next two largest cities (Beijing, Guangzhou) in the second class, then the four (Wuhan, Chongqing, Shenzhen, Tianjin) in the third class.... The city number in each level can be expressed as exponential function such as $f_m=2^{m-1}\approx 0.5e^{-0.693m}$, and the average city size follows an exponential distribution in the form

$$\hat{P}_m = 32511457.634 e^{-0.574m} = 18315332.852 \times 1.775^{1-m}.$$

The goodness of fit is around $R^2=0.993$, and fractal dimension is estimated as $D\approx 0.693/0.574 \approx 1.207$ (See Appendix 2). For this example, the first data point is regarded as an exceptional value, too.

A scaling analysis can be made by using the way similar to the case of the U.S. cities. A least squares calculation involving the data points within the scaling range yields

$$\hat{f}_m = 521201773.262 P_m^{-1.193}.$$

The correlation coefficient square is $R^2\approx 0.997$, and the fractal dimension of China's urban hierarchy is about $D=1.193$ (the standard error $\delta\approx 0.028$). Accordingly, a least squares computation based on equation (4) with the data from rank 3 to 511 (see Figure 3), which corresponds to the scaling range of the urban hierarchy shown in Figure 4(b), results in a model such as

$$P_k = 34568712.629 k^{-0.889}.$$

The goodness of fit is about $R^2=0.995$, and the fractal dimension of city-size distribution is $D=1/q\approx 1/0.889\approx 1.125$, near the value 1.193. This result fails to support the 3/4 power law, and suggests that a single magic number cannot unlock all of nature's secrets (Weibel, 2002).

The American cities and Chinese cities developed in different geographical environments, under different historical and political conditions and backgrounds. However, both the American cities and Chinese cities follow the same hierarchical power laws and exponential laws and take on similar rank-size distributions and cascade structures. First, the cities of both countries comply with Zipf's law. Second, the cities of both countries can be organized into a self-similar hierarchy in light of the generalized $2^n$ rule. Third, the scaling ranges of the hierarchies of cities of both countries are consistent with each other on the whole. Fourth, the fractal dimension values of city size distribution and hierarchical structure are greater than 1. The differences between the two countries' hierarchies of cities are as follows. First, the sizes of the largest cities in America are larger than what is expected in theory (overdeveloped), while the population of the top cities in China is smaller than the predicted values (underdeveloped). Second, the lame-duck class of America's urban hierarchy is due to the absence of city data, while that of China's hierarchy is owing to the undergrowth of small cities and towns (Table 4). Notwithstanding these differences, the cities of both countries lend empirical supports to the claim that Zipf's law formulated by equation (4) is equivalent to the exponential laws expressed with equations (1) and (2), and can be reconstructed as the hierarchical scaling law defined by equation (3).

**Table 4** Comparison of the scaling relations between hierarchy of America's cities and that of China's cities in 2000

| Content | USA cities | PRC cities |
| --- | --- | --- |
| Scaling range | Class 3-class 9 | Class 3-class 9 |
| Corresponding rank | Rank 4-rank 511 | Rank 4-rank 511 |
| Number ratio ($r_f$) | 2 | 2 |
| Average size ratio ($r_p$) | 1.770 | 1.753 |
| Average size ratio based on scaling range ($r_p^*$) | 1.737 | 1.738 |
| Zipf's exponent ($q$) | 0.738 | 0.889 |
| Fractal dimension ($D$) | 1.364 ($\delta \approx 0.021$) | 1.193 ($\delta \approx 0.028$) |
| Cities in top class | Overdeveloped | Underdeveloped |
| Cities in bottom class | Well-developed | Undergrown |

**Note:** The number ratio $r_f=2$ is *ad hoc* given in advance according to theoretical postulate. The average size ratio ($r_p$) is based on the whole values of size ratios, while the adjusted average size ratio ($r_p^*$) is based on the scaling range from the 3rd class to the 9th class.

## 4 Discussion

### 4.1 A theoretical framework of hierarchical scaling and rescaling

Though the hierarchical scaling law is equivalent to Zipf's law, the former seems to be more relevant than the latter. The self-similar hierarchy reveals more information about city development than the Zipf distribution, and it can help solve many problems of urban studies. The analytical process can be illustrated with a 'flow chart', as seen in Figure 5. More importantly, this flow chart can extend in several directions, including allometric scaling, fractals, and $1/f^\beta$ noise. The hierarchy based on the $q$-sequence seems to be a magic framework, from which we can derive many simple scaling rules of cities. Zipf's law (including the rank-size rule), the generalized $2^n$ principle (including Davis' $2^n$ rule), and the hierarchical scaling law are naturally associated with the self-similar hierarchy. Furthermore, we can derive the Pareto distribution and allometric scaling, and the results can be generalized to understanding the $1/f^\beta$ noise.

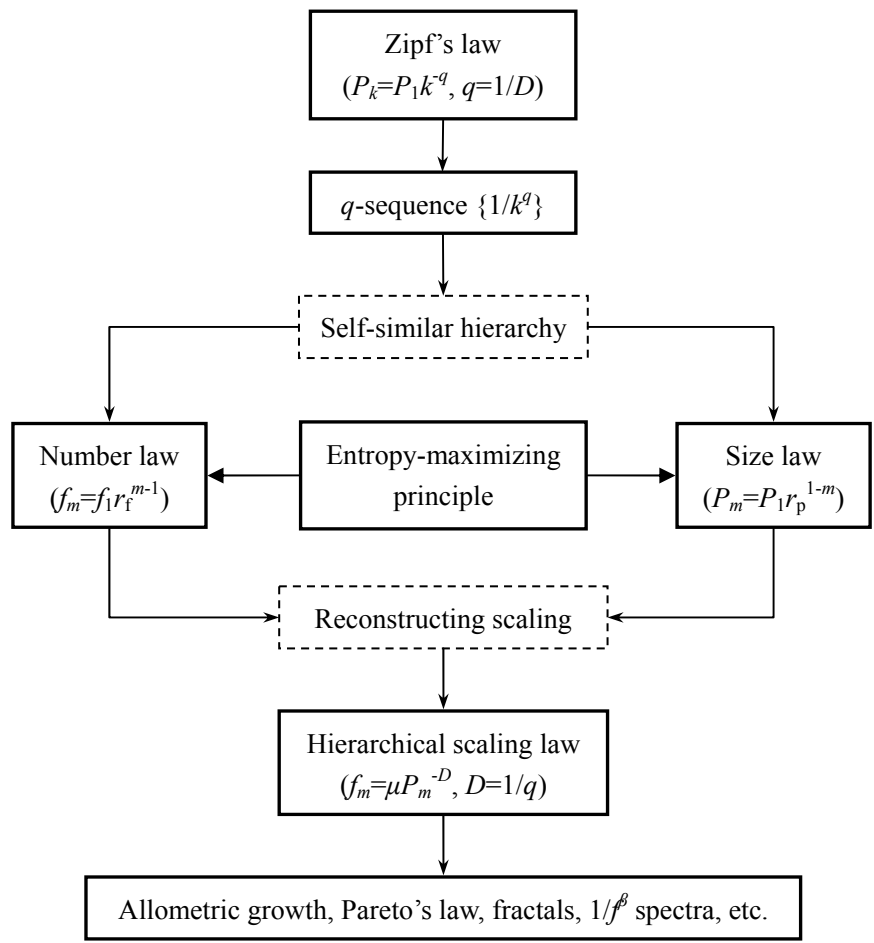

**Figure 5** A logic framework of the laws and rules on city development and urban evolution

First, the Pareto distribution can be derived from the self-similar hierarchy. The hierarchical structure is displayed in Figure 1. The size of the first city in the $m$th class is $P_k=1/(2^{m-1})^q$, where $k=2^{m-1}$. The number of cities with a size greater than $1/(2^{m-1}+1)^q$ is just $2^{m-1}$ (Table 5). Defining a critical size scale $\xi=1/(2^{m-1})^q$, we will have $N(P\geq\xi)=2^{m-1}=\xi^{1/q}$, where $N(P\geq\xi)$ denotes the cumulative number of cities greater than or equal to $\xi$. Generally, the scaling relation can be expressed as

$$N(P \geq \xi) = N_1 \xi^{-1/q} = N_1 \xi^{-D}, \quad (21)$$

where $N_1=1$ denotes the proportionality constant, and $D=1/q$ is regarded as the fractal dimension of the city-size distribution. Obviously, dividing the two sides of equation (21) with the total number of cities gives the formula of Pareto's distribution. This result lends further support to the arguments that the Pareto distribution and Zipf distribution are two different sides of the same coin (Newman, 2005).

Second, the allometric scaling law can be derived from the self-similar hierarchy. The allometric scaling relations can be found in natural and human systems, including urban systems (Batty and Longley, 1994; Bettencourt *et al*, 2007; Chen, 2009; West *et al*, 2002). The generalized allometric scaling law includes various power laws such as Heaps' law (Heaps, 1978; Lü *et al*, 2010; Serrano *et al*, 2009). Empirically, both city population and urban area follow Zipf's law (Chen and Zhou, 2008; Jiang and Liu, 2011). If the population size measure, $P$, is replaced by the urban area measure, $A$, in equation (4), another exponential law can be obtained such as

$$A_m = A_1 r_a^{1-m}, \quad (22)$$

where $A_m$ is the average urban area in the $m$th class, $A_1$ and $r_a$ are parameters, and $r_a=A_m/A_{m+1}$ denotes the interclass *area ratio* (Chen, 2009). An allometric scaling relation between urban area and population can be derived from equations (2) and (22) as follows

$$A_m = a P_m^b, \quad (23)$$

in which $a=A_1 P_1^{-b}$ refers to the proportionality coefficient, and $b=\ln r_a/\ln r_p$ to the scaling exponent.

Third, the results can be generalized to the $1/f^\beta$ noise. The frequency-spectrum relation indicative of the $1/f^\beta$ noise in statistical signal processing and physics can be written as

$$E_f = E_1 f^{-\beta}, \quad (24)$$

where $f=n/T$ denotes frequency ($n=1, \ldots, T/2$, and $T$ is the length of sample path), $E_f$ indicates the corresponding spectral density, $E_1$ and $\beta$ are two parameters, and $\beta$ is what is called spectral exponent (Bak, 1996). Clearly, equation (24) is identical in form to equation (4), and can be abstracted as a $q$-sequence. Maybe we should name it $\beta$-sequence for the random signals. This suggests that the $1/f^{\beta}$ distribution can also be transformed into a hierarchical structure by analogy with the logical procedure displayed in Figures 1 and 5.

Table 5 The results of rearranging the *q*-sequence in terms of the critical size scales based on a geometric sequence

| Classes (*m*) | Rank (*k*) | Size scale ($\xi$) | Elements | Cumulative number $N(P \geq \xi)$ |
|---|---|---|---|---|
| 1--1 | $2^0$ | 1 | 1 | 1 |
| 1--2 | $2^1$ | $1/2^q$ | 1, $1/2^q$ | 2 |
| 1--3 | $2^2$ | $1/4^q$ | 1, $1/2^q$, $1/3^q$, $1/4^q$ | 4 |
| … | … | … | …… | …… |
| 1--*m* | $2^{m-1}$ | $1/(2^{m-1})^q$ | 1, $1/2^q$, $1/3^q$, …, $1/(2^{m-1})^q$ | $2^{m-1}$ |

## 4.2 Theoretical explanations and possibility for generalization

Hierarchy is a form of organization of complex systems which depend on or produce a strong differentiation in capacity (power and size) between the parts of the whole, and it represents a universal structure in both physical and human systems (Chen, 2009; Kleinberg, 2003; Pumain, 2006). The hierarchical scaling behaviors occur in a wide variety of phenomena, including cities, rivers, earthquakes, language, business firms, societies, living species, fractals, and routes from bifurcation to chaos. Now, we have a new way of looking at the relations between the universal ubiquitous empirical observations such as fractals, $1/f$ noise, and Zipf's law. On the surface, no analogy exists between them. In fact, the generality of these phenomena rests with the ubiquity of self-similar hierarchy. First, the $1/f^{\beta}$ noise is analogous to the general form of Zipf's law, and $1/f$ noise is a special case of the $1/f^{\beta}$ noise. Both the $1/f$ distribution and the Zipf distribution can be converted into a hierarchical structure. Second, a fractal system is also a hierarchy with a cascade structure, and the $1/f$ noise and Zipf's law indicates some kind of fractality. Third, the power laws of fractals, $1/f$ noise, and Zipf's law can be converted into two exponential laws. The exponential laws suggest the most probable distributions, which can be derived by using the entropy-maximizing methods (Wilson, 2000). There have been varied and interesting explanations and interpretations regarding Zipf's law and related scaling laws (Bettencourt *et al*, 2007; Cancho and Solé, 2003; Carlson and Doyle, 1999; Ebeling and Pöschel, 1994; Ferrer i Cancho and Solé, 2003; Ferrer i Cancho *et al*, 2005; Kanter and Kessler, 1995; Marsili and Zhang, 1996; Podobnik *et al*, 2010). Recent research shows that Zipf's law can be derived from entropy-maximizing postulates (Chen, 2012).

The important knowledge fields beneficial to our understanding of how cities evolve include fractal geometry, allometric growth, and network science (Batty, 2008). The transformative relationship between Zipf's law and the generalized $2^n$ rule is useful to reveal the similarities and differences between these theories. An interesting discovery is that, if the scaling exponent is $q$=1, the generalized $2^n$ principle can be further generalized to the $3^n$ principle, the $4^n$ principle, or even the $N^n$ principle ($N$=2, 3, 4, …) (This will be discussed in a companion paper). The results can be expanded to describe river networks, energy distribution of earthquakes, etc. Because of the translational symmetry of the exponential distribution, if the top-down order of classes is changed into a bottom-up order, the model's structure will not vary. Thus, equations (1) and (2) are identical in mathematical expressions to Horton-Strahler's laws in geomorphology (Horton, 1945; Strahler, 1952) and Gutenberg-Richter's laws in seismology (Gutenberg and Richter, 1954). This implies that exponential laws and related power laws can be used to characterize various similar networks and hierarchies in the physical world and in human systems.

Several new problems remain to be solved in the future. First, if $q$=1 as given, equation (4) can be converted into the $2^n$ rule ($r_f$=$r_p$=2), $3^n$ rule ($r_f$=$r_p$=3), $4^n$ rule ($r_f$=$r_p$=4), and so on. In theory, the one-parameter Zipf's formula $P_k$=$P_1$/$k$ is equivalent to the $N^n$ rule ($r_f$=$r_p$=$N$). However, if $q$≠1, the two-parameter Zipf's law seems to be only converted into the generalized $2^n$ principle ($r_f$=2, $r_p$=$2^q$). Second, if an inverse transform is implemented from equations (1) and (2) to equation (4), a three-parameter Zipf's model will be obtained (Chen and Zhou, 2003; Mandelbrot, 1983). This seems to suggest that the largest city is independent of the scaling law--number one is a special one. This also suggests that there is something unclear about the relationships between Zipf's law and the hierarchical scaling law.

## 5 Conclusions

In this paper, I define a self-similar hierarchy of cities based on the *q*-sequence abstracted from Zipf's law. One of the main academic contributions of this work to science is in deriving two exponential laws from the hierarchy of the *q*-sequence, and the two exponential laws imply the hierarchical scaling law. Thus, in this way, Zipf's law can be transformed into the hierarchical scaling law. Though the mathematical proof suggests an equivalent relationship between Zipf's law and the hierarchical scaling law, the latter is better for describing urban systems than the former. First, the hierarchical scaling law shows a clear mathematical structure. This structure bears an analogy with fractal recurrence. Second, the hierarchical scaling law has clear physical meaning. It can be derived by using the method of entropy-maximization. Third, the hierarchical scaling law can be associated with many mathematical rules of cities. It can be generalized to encompass the $2^n$ rule, Pareto's law, the allometric growth law, and the $1/f^\beta$ spectra, etc. Additionally, it shares many similarities with Horton-Strahler's laws in geomorphology and Gutenberg-Richter's laws in seismology.

The main findings and conclusions of this research are as follows. First, Zipf's law is a signature of the hierarchical scaling. Zipf distributions often suggest hierarchical structure. We can use the rank-size scaling to search for self-similar hierarchy. Second, the self-similar hierarchy can be used to rescale Zipf's law. According to the mathematical proof in this article, Zipf's distribution can always be transformed into a hierarchy with a cascade structure. By doing so, we can reveal useful information about urban evolution. Third, the scaling exponent determines the common ratio of the self-similar hierarchy. If the exponent is unequal to 1, the number ratio of cities must be 2, and Zipf's law suggests the generalized $2^n$ rule. However, if the exponent equals 1, the number ratios can be an arbitrary positive integer such as 2, 3, 4, 5, and 6. Fourth, a new network theory may be developed by means of hierarchical scaling. The hierarchy and network are actually different sides of the same coin (Batty and Longley, 1994). From the self-similar hierarchical structure, we can derive a fractal network indicating recursive subdivision of space.

## Acknowledgements:

This research was sponsored by the National Natural Science Foundation of China (Grant No. 41171129). The support is gratefully acknowledged. I would also like to thank three anonymous reviewers whose interesting comments were helpful in improving the quality of this paper.

# References

Altmann EG, Pierrehumbert JB, Motter AE (2009). Beyond word frequency: Bursts, lulls, and scaling in the temporal distributions of words. *PLoS ONE*, 4(11): e7678

Axtell RL (2001). Zipf distribution of U.S. firm sizes. *Science*, 293: 1818-1820

Bak P (1996). *How Nature Works: the Science of Self-Organized Criticality*. New York: Springer-Verlag

Basu B, Bandyapadhyay S (2009). Zipf's law and distribution of population in Indian cities. *Indian Journal of Physics*, 83(11): 1575-1582

Batty M (2006). Rank clocks. *Nature*, 444: 592-596

Batty M (2008). The size, scale, and shape of cities. *Science*, 319: 769-771

Batty M, Longley PA (1994). *Fractal Cities: A Geometry of Form and Function*. London: Academic Press

Bettencourt LMA, Lobo J, Helbing D, Kühnert C, West GB (2007). Growth, innovation, scaling, and the pace of life in cities. *PNAS*, 104 (17): 7301-7306

Blasius B, Tönjes R (2009). Zipf's law in the popularity distribution of chess openings. *Physical Review Letters*, 103(21): 218701

Brakman S, Garretsen H, Van Marrewijk C, Van Den Berg M (1999). The return of Zipf: Towards a further understanding of the rank-size distribution. *Journal of Regional Science,* 39(1): 183–213

Cancho RFi, Solé RV (2003). Least effort and the origins of scaling in human language. *PNAS*, 100(3): 788-791

Carlson JM, Doyle J (1999). Highly optimized tolerance: A mechanism for power laws in designed systems. *Physical Review E*, 60(2): 1412-1427

Carroll C (1982). National city-size distributions: What do we know after 67 years of research? *Progress in Human Geography*, 6(1): 1-43

Chen YG (2009). Analogies between urban hierarchies and river networks: Fractals, symmetry, and self-organized criticality. *Chaos, Soliton & Fractals*, 40(4): 1766-1778

Chen YG (2011). Modeling fractal structure of city-size distributions using correlation functions. *PLoS ONE*, 6(9):e24791

Chen YG (2012). The rank-size scaling law and entropy-maximizing principle. *Physica A: Statistical*

*Mechanics and its Applications*, 391(3): 767-778

Chen YG, Zhou YX (2003). The rank-size rule and fractal hierarchies of cities: mathematical models and empirical analyses. *Environment and Planning B: Planning and Design*, 30(6): 799-818

Chen YG, Zhou YX (2008). Scaling laws and indications of self-organized criticality in urban systems. *Chaos, Soliton & Fractals*, 35(1): 85-98

Clark C (1951). Urban population densities. *Journal of Royal Statistical Society*, 114(4): 490-496

Córdoba, J-C (2008). On the distribution of city sizes. *Journal of Urban Economics*, 63(1): 177-197

Davis K (1978). World urbanization: 1950-1970. In: *Systems of Cities*. Eds. I.S. Bourne and J.W. Simons. New York: Oxford University Press, pp92-100

Ebeling W, Pöschel T (1994). Entropy and long range correlations in literary English. *Europhysics Letters*, 26(4): 241–246

Ferrer i Cancho R, Riordan O, Bollobás B (2005). The consequences of Zipf's law for syntax and symbolic reference. *Proceedings of the Royal Society B: Biological Sciences*, 272(1562): 561-565

Ferrer i Cancho R, Solé RV (2003). Least effort and the origins of scaling in human language. *PNAS*, 100(3): 788–791

Flam F (1994). Hints of a language in junk DNA. *Science*, 266: 1320

Frankhauser P (1990). Aspects fractals des structures urbaines. *L'Espace Géographique*, 19(1): 45-69

Frankhauser P (1998). The fractal approach: A new tool for the spatial Analysis of urban agglomerations. *Population: An English Selection*, 10(1): 205-240

Furusawa C, Kaneko K (2003). Zipf's law in gene expression. *Physical Review Letters*, 90(8): 088102

Gabaix X (1999). Zipf's law for cities: an explanation. *Quarterly Journal of Economics*, 114 (3): 739–767

Gabaix X (2009). Power laws in economics and finance. *The Annual Review of Economics*, 1:255-593

Gabaix X, Ioannides YM (2004). The evolution of city size distributions. In: *Handbook of Urban and Regional Economics, Vol. 4*. Eds. J. V. Henderson and J. F. Thisse. Amsterdam: North-Holland Publishing Company, pp 2341-2378

Gutenberg B, Richter CF (1954). *Seismicity of the Earth and Associated Phenomenon* (2nd ed.). Princeton: Princeton University Press

Heaps HS (1978). *Information Retrieval: Computational and Theoretical Aspects*. Orlando: Academic Press

Holland J (1995). *Hidden Order: How Adaptation Builds Complexity*. Reading, MA: Addison-Wesley

Horton RE (1945). Erosional development of streams and their drainage basins: hydrophysical approach to quantitative morphology. *Bulletin of the Geophysical Society of America*, 56(3): 275-370

Jiang B, Liu X (2011). Scaling of geographic space from the perspective of city and field blocks and using volunteered geographic information. *International Journal of Geographical Information Science*, *arXiv*:1009.3635v4 [physics.data-an]

Jiang B, Yao X. (2010 Eds.). *Geospatial Analysis and Modeling of Urban Structure and Dynamics*. New York: Springer-Verlag

Jullien R, Botet R (1987). *Aggregation and Fractal Aggregates*. Singapore: World Scientific Publishing Co.

Kanter I, Kessler DA (1995). Markov processes: Linguistics and Zipf's law. *Physical Review Letters*, 74(22): 4559-4562

Kleinberg J (2003). Bursty and hierarchical structure in streams. *Data Mining and Knowledge Discovery*, 7(4): 373-397

Knox PL, Marston SA (2006). *Places and Regions in Global Context: Human Geography (4th Edition)*. Upper Saddle River, NJ: Prentice Hall

Lü LY, Zhang ZK, Zhou T (2010). Zipf's law leads to Heaps' law: analyzing their relation in finite-size systems. *PLoS ONE*, 5(12): e14139

Makse HA, Andrade Jr. JS, Batty M, Havlin S, Stanley HE (1998). Modeling urban growth patterns with correlated percolation. *Physical Review E*, 58(6): 7054-7062

Makse HA, Havlin S, Stanley HE (1995). Modelling urban growth patterns. *Nature*, 377: 608-612

Mandelbrot BB (1983). *The Fractal Geometry of Nature*. New York: W. H. Freeman and Company

Manrubia S, Zanette D (1998). Intermittency model for urban development. *Physical Review E*, 58(1): 295-302

Marsili M, Zhang YC (1996). Interacting individuals leading to Zipf's law. *Physical Review Letters*, 80(12): 2741–2744

Newman MEJ (2005) Power laws, Pareto distributions and Zipf's law. *Contemporary Physics*, 46(5): 323–351

Petersen AM, Podobnik B, Horvatic D, Stanley HE (2010). Scale-invariant properties of public-debt

growth. *Europhysics Letters*, 90, 38006

Podobnik B, Horvatic D, Petersen AM, Urošević B, Stanley HE (2010). Bankruptcy risk model and empirical tests. *PNAS*, 107(43): 18325-18330

Pumain D (2006 Ed.). *Hierarchy in Natural and Social Sciences*. Dordrecht: Springer

Rozenfeld HD, Rybski D, Andrade Jr. JS, Batty M, Stanley HE, Makse HA (2008). Laws of population growth. *PNAS*, 105(48): 18702-18707

Rozenfeld HD, Rybski D, Gabaix X, Makse HA (2011). The area and population of cities: New insights from a different perspective on cities. *American Economic Review*, 101(5): 2205-2225

Serrano MÁ, Flammini A, Menczer F (2009). Modeling statistical properties of written text. *PLoS ONE*, 4: e5372

Shao J, Ivanov PCh, Podobnik B, Stanley HE (2007). Quantitative relations between corruption and economic factors. *The European Physical Journal B-Condensed Matter and Complex Systems*, 56(2): 157-166

Shao J, Ivanov PCh, Urošević B, Stanley HE, Podobnik B (2011). Zipf rank approach and cross-country convergence of incomes. *Europhysics Letters*, 94, 48001

Stanley MHR, Buldyrev SV, Havlin S, Mantegna RN, Salinger MA, Stanley HE (1995). Zipf plots and the size distribution of firms. *Economics Letters*, 49(4): 453-457

Strahler AE (1952). Hypsometric (area-altitude) analysis of erosional topography. *Geological Society of American Bulletin*, 63(11): 1117-1142

Vicsek T (1989). *Fractal Growth Phenomena*. Singapore: World Scientific Publishing Co.

Weibel ER (2002). The pitfalls of power laws. *Nature*, 417: 131-132

West GB, Brown JH, Enquist BJ (1999). The fourth dimension of life: fractal geometry and allometric scaling of organisms. *Science*, 284: 1677–1679

West GB, Woodruff WH, Brown JH (2002). Allometric scaling of metabolic rate from molecules and mitochondria to cells and mammals. *PNAS*, 99(s11): 2473-2478

Wilson AG (2000). *Complex Spatial Systems: The Modelling Foundations of Urban and Regional Analysis*. Singapore: Pearson Education

Zanette D, Manrubia S (1997). Role of intermittency in urban development: a model of large-scale city formation. *Physical Review Letters*, 79(3): 523-526

Zipf GK (1949). *Human Behavior and the Principle of Least Effort*. Reading, MA: Addison-Wesley

# Appendices

## 1. A self-similar hierarchy in mathematics

A fractal is a typical hierarchy with a cascade structure. Let's see the well-known growing fractal presented by Jullien and Botet (1987) and discussed by Vicsek (1989). In the first class (level), there is one fractal copy (the number is $N_1=1$), and the linear size of the copy can be regarded as one unit (the 'side length' of the fractal copy is $L_1=1$); In the second class, there are five fractal copies (the number is $N_2=5$), and the linear size of the copies is one third (the 'length' of the fractal copies is $L_2=1/3$). Generally, in the $m$th class, the number of the fractal copies is $N_m=5^{m-1}$, and the linear size of the copies is $L_m=3^{1-m}$ (Figure A). The hierarchical structure of the growing fractal can be described with equations (1) and (2), which yield

$$N_m = N_1 r_n^{m-1} = 5^{m-1}, \tag{A1}$$

$$L_m = L_1 r_l^{1-m} = 3^{1-m}, \tag{A2}$$

where the common ratios, $r_n$ and $r_l$, are as below:

$$r_n = N_{m+1}/N_m = 5, \quad r_l = L_m/L_{m+1} = 3.$$

From equations (A1) and (A2) follows a scaling relation such as:

$$N_m = \zeta L_m^{-D}, \tag{A3}$$

where the proportionality coefficient $\zeta=N_1 L_1^D$, and the parameter $D$ denotes fractal dimension, which can be defined by

$$D = \frac{\ln r_n}{\ln r_l} = \frac{\ln(5)}{\ln(3)} \approx 1.465. \tag{A4}$$

It can be generalized from the above mathematical process to other regular fractals and statistical fractals. The simple regular fractals do not follow Zipf's law, but the random fractals and multifractals may comply with the general form of Zipf's law.

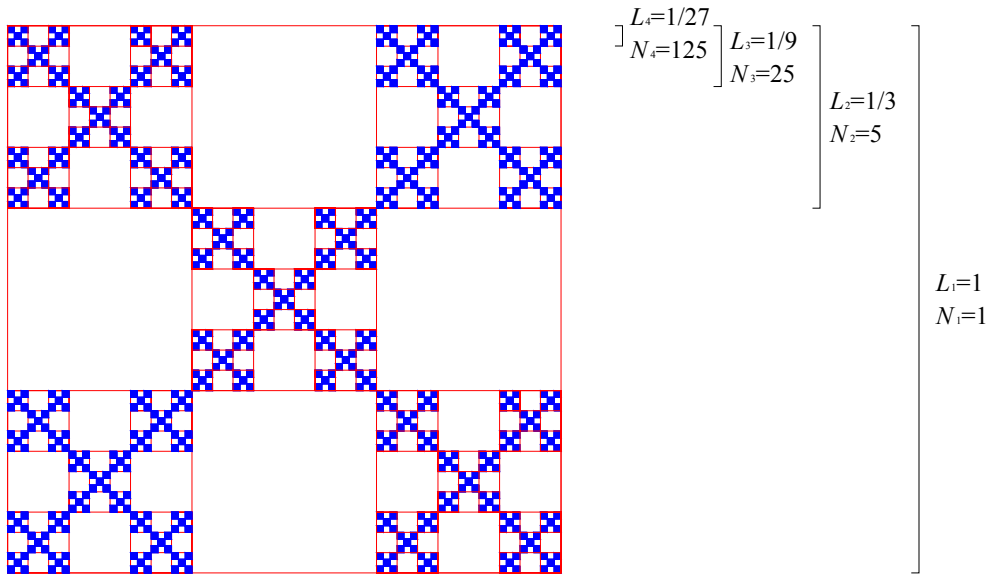

Figure A sketch map of the hierarchy of a growing fractal (only the first 5 levels are shown)

## 2. The exponential distributions of average city sizes in hierarchies of cities

In the empirical analyses, the hierarchies of cities are constructed in terms of equation (1), and the number ratio is taken as $r_f$=2. Whether or not the city size and number follow the hierarchical scaling law, the key is to see whether or not the average city sizes of different levels satisfy the exponential distribution and can be fitted into equation (2). The answer is 'yes' for both American cities and Chinese cities in 2000, and the semi-logarithmic plots are displayed in Figure B.

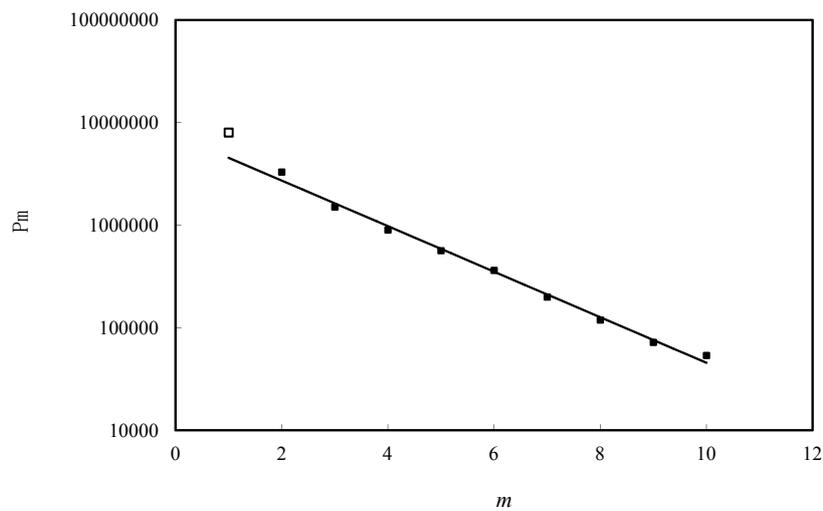

a. USA cities

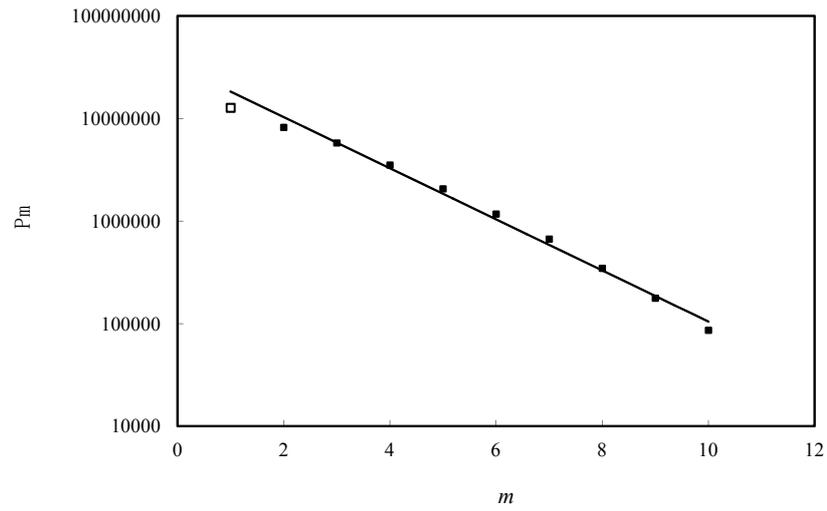

b. PRC cities

**Figure B** The exponential distributions of average sizes in the hierarchy of American cities and that of

Chinese cities (2000)

(**Note:** According to Clark (1951), the first data points are treated as exceptional values and not taken into the least squares calculations.)